\documentclass{article}
\usepackage{sw20nhj}
\usepackage{amsmath}
\usepackage{amsfonts}
\usepackage{amssymb}
\usepackage{graphicx}

\begin{document}
\author{Vladimir K. Petrov\thanks{
E-mail address: petrov@earthling.net}}
\title{{\Large POSSIBLE CONSEQUENCES OF CONJECTURAL PERIODICITY OF SPECTRUM OF
LATTICE DIRAC OPERATOR}}
\date{{\it N. N. Bogolyubov Institute for Theoretical Physics}\\
{\it \ National Academy of Sciences of Ukraine}\\
{\it \ 252143 Kiev, Ukraine. 5.25.1998}}
\maketitle
\begin{abstract}
Some consequences which follow from the periodicity assumption for spectral
density of Wilson--Dirac operator are studied. Such an assumption allows to
obtain simple representations for quark propagator, which reveals an
important role of $m\leftrightarrow -m$ symmetry. It is argued that this
symmetry is restored when the mirror fermion mass $m_r$ tended to infinity.
The constrains on zero modes of Wilson--Dirac operator in a toy model
approximation are also discussed.
\end{abstract}

\section{Introduction}

The spectrum of Dirac operator is closely related to a number of important
aspects in finite temperature QCD. The behavior of spectral density
$\rho(\lambda)$ around $\lambda\sim0$ for massless Dirac operator is of
particular physical interest being responsible for breaking the chiral
symmetry \cite{BC}. The low lying eigenvalues of staggered fermion operator
are also used to extract the topological content of background gauge field
configurations \cite{EHN}.

Despite steady progress, the numerical simulation results for $\rho(\lambda)$
remain inconclusive. As it is known from general theoretical considerations
(see e.g. \cite{Lu,BJJLSS,JLSS}), the eigenvalues of the normalized Dirac
operator are located in a compact area. However, as it pointed out in~\cite{S}
perturbative calculations show even an increase of state density at large
$\lambda$. Indeed, for $\lambda\gg\Lambda_{QCD}$ the spectral density is
regarded to be insensitive to gluon vacuum fluctuations and behaves in the
same way as for free fermions:
\begin{equation}
\rho^{free}(\lambda)\sim\frac{N_{c}\lambda^{3}}{4\pi^{2}}.
\end{equation}
This free term introduces quadratically divergent piece in $\langle\bar{\psi
}\psi\rangle$. However, to get a truly non-perturbative quark condensate which
is the order parameter of the spontaneous chiral symmetry breaking, this
perturbative divergent part should be subtracted. As a result, the quark
condensate is related to the region of small $\lambda$: $\lambda\sim
m\ll\Lambda_{QCD}$ \cite{S}.

Some numerical results \cite{K,K1,K2} may be taken as an evidence in favor of
compactness of Dirac operator spectrum (see Fig.~2,4,5 and 6 of ref.~\cite{K}%
). Indeed, the spectral density disappears outside certain region
$\Gamma_{\lambda}$ \cite{OV}, that for a particular case may be specified as:
$\left|  \lambda\right|  \lesssim2\Lambda_{QCD}$ (see Fig.~2 of ref.~\cite{S}%
), or for $\left|  \lambda\right|  \lesssim3/C$ (see Fig.~2 of ref.~\cite{St}%
). The domain wall fermion approach leads to a similar picture (see \cite{J}
Fig. 12). In case of (1+1)-dimension it was proved \cite{GHL,BDET,BDET2}, that
the spectrum was contained inside the compact area (circle of radius 2 in the
complex plane). Moreover, compactness more often than not may be treated as
implicit periodicity, so it is hard to avoid the assumption that the spectral
density may appear a periodic function of $\lambda.$ Indeed, the lattice
regularization for Dirac operator leads to
\begin{equation}
D\left(  A\right)  =\;{\gamma}^{\nu}(\partial_{\nu}\;-\;iA_{\nu}%
)-m\rightarrow\sum_{\nu=0}^{3}D\left(  A\right)  _{x^{\prime}x}^{\nu},
\end{equation}
where
\begin{equation}
D\left(  A\right)  _{x^{\prime}x}^{\nu}\ =\frac{r-\gamma_{\nu}}{2}%
\!\!\!U_{\nu}\left(  x\right)  \delta_{x,x^{\prime}-\nu}+\frac{r+\gamma_{\nu}%
}{2}U_{\nu}^{\dagger}\left(  x^{\prime}\right)  \delta_{x,x^{\prime}+\nu
}-\left(  r+\delta_{\nu}^{0}ma\right)  \delta_{x^{\prime},x}\label{D}%
\end{equation}
and $r\equiv\tanh a_{\tau}m_{r}$ Wilson parameter. Let us assume that the
fermion part of the action $-S_{F}=\ln\det D$ can be expressed in terms of
Polyakov loops $\Omega\left(  \mathbf{x}\right)  =\prod_{t=0}^{N_{\tau}%
-1}U_{0}\left(  t,\mathbf{x}\right)  ,$ as it can be done in numerous
particular cases for the gluon part of the action, and $\det D=f\left(
\varphi\right)  $ . In this case one may expect that the Dirac matrix $\det
D\left(  \varphi\right)  $ (as well as $\left\langle {\bar{\psi}\psi
}\right\rangle $) became periodic in $\varphi$
\begin{equation}
\det D\left(  A_{0}\right)  =\det D\left(  A_{0}+2\pi T\right)  ,
\end{equation}
with
\begin{equation}
\varphi=\arg\Omega=a_{\tau}N_{\tau}A_{0}\equiv\frac{A_{0}}{T}.
\end{equation}
Therefore, one may suspect that this condition, together with the periodicity
along $t$ - axis, may induce the periodicity of $D$ in the complex plane of
$m$ and, hence, in eigenvalues $\lambda.$ In other words it would not be too
surprising if
\begin{equation}
\lambda_{n}\left(  \varphi+2\pi\right)  =\lambda_{n}\left(  \varphi\right)
+2\pi\Theta;\quad\Theta=const.\label{Ds}%
\end{equation}

Indeed, if to take the free Dirac operator as an example \cite{St}, the
spectrum in the phase $\arg\Omega=0$ is given by: $\lambda^{2}=\mathbf{k}%
^{2}+((2n+1)\pi T)^{2}$ and $\lambda^{2}=\mathbf{k}^{2}+((2n+1/3)\pi T)^{2}$
for $\arg\Omega=2\pi/3$.

More rigorous and convincing evidence in favor of the periodicity of Dirac
operator spectrum in general case was suggested in \cite{Ne98} (see also
\cite{Na,SJ}). Indeed, Dirac operator can be presented \cite{Ne} as
\begin{equation}
aD=1-\gamma_{5}\epsilon(H)),\label{d4}%
\end{equation}
where $\epsilon(H)$ some Hermitian matrix and $H=\gamma_{5}[1-\gamma_{\mu
}(\partial_{\mu}+iA_{\mu})]$. Therefore, if $D$\ satisfies the Ginsparg-Wilson
relation \cite{GW}
\begin{equation}
D\gamma_{5}+\gamma_{5}D=aD\gamma_{5}D,
\end{equation}
then the operator $\gamma_{5}\epsilon(H)=1-aD$\ is unitary
\begin{equation}
\left(  1-aD\ \right)  ^{\dagger}\left(  1-aD\ \right)  =\left(  1-a\gamma
_{5}D\gamma_{5}\ \right)  \left(  1-aD\ \right)  =1.
\end{equation}
This implies that the spectrum of $D$\ lies in the complex plane
\cite{Ne98,Na,SJ} on the circle, $(1-e^{i\theta})/a$\ with $\theta\in[-\pi
,\pi]$\textrm{.} Considering the above, a Banks-Casher-like relation \cite{BC}
will be
\begin{equation}
\langle\bar\psi\psi\rangle=\langle D^{-1}\left(  x,x\right)  \rangle\sim
a\int_{-\pi}^{\pi}\frac{\rho\left(  \theta\right)  }{1+am-e^{i\theta}}%
d\theta.\label{ne}%
\end{equation}

The periodicity of Dirac spectrum $\rho\left(  \theta\right)  $ in $\theta$
may be, as a matter of course, treated as a lattice artefact Indeed, assuming
that the major part of eigenstates is concentrated in the area $\theta
^{2}\lesssim a^{2},$ \cite{SJ} after redefinition of the integration variable
$\theta\rightarrow sa,$ and lifting lattice regularization $\left(
a\rightarrow0\right)  $ one comes to the standard Banks-Casher relation
\begin{equation}
\langle\bar\psi\psi\rangle=\langle D^{-1}\left(  x,x\right)  \rangle\sim
\int_{-\frac\pi a}^{\frac\pi a}\frac{\rho\left(  sa\right)  }{\frac{1-e^{isa}%
}a+m}ds\simeq\int_{-\infty}^{\infty}\frac{\rho\left(  sa\right)  }%
{m-is}ds.\label{SJ}%
\end{equation}

Though, one should also consider the possibility that e.g., $\lambda
^{2}\lesssim1/T^{2}$, with $\lambda\equiv\theta T.$ In this case the spectrum
remains invariant under the shift $\lambda\rightarrow\lambda\pm2\pi T.$ The
Banks-Casher relation, of course, remains valid in all cases, however,
premature lifting the lattice regularization in $\left(  \ref{SJ}\right)  ,$
leaves the conjectural periodicity hidden.

Certainly, the arguments given above, will not suffice as such and we must
admit, that the assumption $\left(  \ref{Ds}\right)  $ looks much more
controversial then the evidences given in favor of compactness. Below we
briefly sketch the results \cite{P} obtained in the approximation, where all
terms proportional to $1/\xi$ have been discarded\footnote{The anisotropy
parameter $\xi$ is defined as $\xi\simeq a_{\sigma}/a_{\tau},$ where $a_{\tau
}$ and $a_{\sigma}$ are temporal and spatial lattice spacings, respectively.}
in a Hamiltonian limit $a_{\tau}\rightarrow0$ $\left(  \xi\rightarrow
\infty\right)  .$ The spectral density computed in such an approximation is
actually periodic. This reassures us a bit and allows us to assume that the
periodicity may appear to be a general feature of the spectral density of
Dirac operator. Hereafter we consider some consequences which follow from such
an assumption for a more general case.

\section{Toy model approximation}

In a toy model approximation the fermion determinant can be computed
analytically and presented in a closed form. We choose the form suggested in
\cite{HK} (see also \cite{BRWS}) for the fermion part of action
\begin{equation}
-S\!\!\!_{F}=n_{f}\left(  a^{3}\sum_{x}\overline{\!\!\!\psi}_{x^{\prime}%
}D_{x^{\prime}x}^{0}\psi_{x}+\frac{a^{3}}\xi\sum_{n=1}^{d}\sum_{x}%
\overline{\!\!\!\psi}_{x^{\prime}}D_{x^{\prime}x}^{n}\psi_{x}\right)
,\label{sf}%
\end{equation}
where $n_{f}$\ refers to the number of flavors\footnote{We consider degenerate
quarks.} and $D_{x^{\prime}x}^{\nu}$ are given by $\left(  \ref{D}\right)  $.

Inasmuch as in the suggested approximation we omit the terms of $1/\xi$ order,
the 'toy' action will be simply $a^{3}\sum_{x}\overline{\!\!\!\psi}%
_{x}D_{xx^{\prime}}^{0}\psi_{x^{\prime}}$. The projectors $\frac{1\pm
\gamma_{0}}2$ divide the bispinors $\psi$ into two components $\psi^{\left(
\pm\right)  }=\frac{1\pm\gamma_{0}}2\psi,$ each including only one
two-component spinor. Therefore, taking into account that $\left(
\delta_{x-\hat\mu,x^{\prime}}\right)  ^{\dagger}=\delta_{x,x-\hat\mu^{\prime}%
}=\delta_{x+\hat\mu,x^{\prime}},$\ and presenting
\begin{equation}
\frac{r\pm\gamma_{0}}2=\frac{1+\gamma_{0}}2\frac{r\pm1}2+\frac{r\mp1}%
2\frac{1-\gamma_{0}}2,
\end{equation}
we may rewrite $\left(  \ref{sf}\right)  $\ as
\begin{equation}
-S_{F}=\sqrt{1-r^{2}}\left(  \bar\psi_{x^{\prime}}^{\left(  +\right)  }%
\Delta_{x^{\prime}x}^{\dagger}\psi_{x}^{\left(  +\right)  }+\bar
\psi_{x^{\prime}}^{\left(  -\right)  }\Delta_{x^{\prime}x}\psi_{x}^{\left(
-\right)  }\right)  ,\label{SF+}%
\end{equation}
where
\begin{equation}
\Delta_{xx^{\prime}}=\delta_{\mathbf{xx}^{\prime}}\left(  \frac{e^{a_{\tau
}m_{r}}U_{0}\left(  \mathbf{x},t\right)  }2\delta_{t^{\prime},t-1}%
-\frac{e^{-a_{\tau}m_{r}}U_{0}^{\dagger}\left(  \mathbf{x},t^{\prime}\right)
}2\delta_{t^{\prime},t+1}-\frac{ma_{\tau}+r}{\sqrt{1-r^{2}}}\delta_{t^{\prime
},t}\right)  .\label{del}%
\end{equation}

By gauge transformation all $U_{0}\left(  \mathbf{x},t\right)  _{\alpha\nu}$
matrices may be diagonalized simultaneously: $U_{0}\left(  \mathbf{x}%
,t\right)  _{\alpha\nu}=\delta_{\alpha\nu}U_{0}\left(  \mathbf{x},t\right)
_{\alpha\alpha},$ therefore $\Delta_{xx^{\prime}}$ in $\left(  \ref{del}%
\right)  $ is simply a set of $N$ matrices $N_{\tau}\times N_{\tau}.$
Hopefully, the computation of $\det\Delta,$ in the considered extremely simple
case can be done straightforwardly and after the integration over $\psi
_{x}^{\left(  \pm\right)  }$ we get
\begin{equation}
-S_{F}^{eff}=n_{f}\sum_{\alpha}\ln\det\Delta_{\alpha}^{\dagger}\Delta_{\alpha
}=n_{f}\ln\prod_{\alpha=1}^{N}\left(  z_{r}-\cos\varphi_{\alpha}\right)
\left(  z-\left(  -1\right)  ^{N_{\tau}}\cos\varphi_{\alpha}\right)
,\label{Seff}%
\end{equation}
with
\begin{equation}
z=\left(  -1\right)  ^{B}\cosh\frac{m}{T};\quad z_{r}=\left(  -1\right)
^{B}\cosh\frac{m+2m_{r}}{T},\label{z}%
\end{equation}
where $B=0$ for periodic and $B=1$ for antiperiodic border conditions on
fermion fields in a temporal direction. Thereafter we shall consider the
standard case $B=1$ and even values $N_{\tau}$, because the generalization on
$B=0$ and odd $N_{\tau}$ is quite evident.

It easy to see from $\left(  \ref{Seff}\right)  $ and $\left(  \ref{z}\right)
$ that ${\langle\,\bar\psi\psi\rangle}$ may be presented as
\begin{align}
{\langle\,\bar\psi\psi\rangle}  & =2n_{f}\sinh\frac mT\left\langle
\frac1{\cosh\frac mT+\cos\varphi}\right\rangle \label{p}\\
& \ \ \ \ \ +2n_{f}\sinh\frac{m+2m_{r}}T\left\langle \frac1{\cosh
\frac{m+2m_{r}}T+\cos\varphi}\right\rangle ,\nonumber
\end{align}
where $m_{r}$ is the mirror fermion mass connected to Wilson parameter by the
relation $a_{\tau}m_{r}=\operatorname{arctanh}r.$

\section{General case}

Recall that the fermion Green's function in the external gauge
field\textrm{\ }is given by
\begin{equation}
\langle\bar{\psi}(y)\psi(x)\rangle=\sum_{n}\frac{u_{n}^{\dagger}(y)u_{n}%
(x)}{m-i\lambda_{n}},\label{ps}%
\end{equation}
where $\lambda_{n}(x)$ eigenvalues of the \textit{massless} Dirac operator and
$u_{n}$ are corresponding eigenfunctions. The spectrum of the massless Dirac
operator is discrete and enjoys\ the chiral symmetry. In particular, for any
eigenfunction $u_{n}(x)$, the function
\begin{equation}
\tilde{u}_{n}=\gamma^{5}u_{n}\label{s1}%
\end{equation}
\ is the eigenfunction as well, with the eigenvalue
\begin{equation}
\tilde{\lambda}_{n}=-\lambda_{n}.\label{ev}%
\end{equation}
Therefore
\begin{equation}
\rho\left(  \lambda\right)  =\rho\left(  -\lambda\right) \label{s2}%
\end{equation}
and eigenfunctions occur in pairs with opposite eigenvalues, except for zero
modes\footnote{Generally, symmetry $\left(  \ref{s2}\right)  $ does not mean
chiral invariance. Indeed, the eigen functions of Dirac operator are the same
for $m=0$ and $m\neq0.$ Therefore, the symmetry $\left(  \ref{s2}\right)  $ is
preserved for arbitrary $m$, though only massless Dirac operator is invariant
under transformation $\left(  \ref{s1}\right)  $.}. In the limit
$V\rightarrow\infty$, the level spectrum becomes dense and we can trade the
sum in $\left(  \ref{ps}\right)  $ for the integral and get
\begin{equation}
\left\langle {\bar{\psi}\psi}\right\rangle =2m\int_{0}^{\infty}\frac
{\rho\left(  \lambda\right)  }{m^{2}+\lambda^{2}}d\lambda,\label{den}%
\end{equation}
where
\begin{equation}
\rho(\lambda)=\frac{1}{V}\left\langle \sum_{n}\delta(\lambda-\lambda
_{n})\right\rangle .\label{rod}%
\end{equation}
In case when spectral density is compact and periodic with the some period
$\Theta$%
\begin{equation}
\rho\left(  \lambda\right)  =\rho\left(  \lambda+2k\pi\Theta\right)
\end{equation}
we can write
\begin{equation}
\left\langle {\bar{\psi}\psi}\right\rangle _{m}=m\sum_{k=-\infty}^{\infty}%
\int_{2\pi k}^{2\pi\left(  k+1\right)  }\frac{\rho\left(  \lambda\right)
}{m^{2}+\lambda^{2}}d\lambda
\end{equation}
or
\begin{equation}
\left\langle {\bar{\psi}\psi}\right\rangle _{m}=m\sum_{k=-\infty}^{\infty}%
\int_{0}^{2\pi}\frac{\rho\left(  \lambda\right)  }{m^{2}+\left(  \lambda+2\pi
k\Theta\right)  ^{2}}d\lambda.\label{s-d}%
\end{equation}
Presenting (\cite{PBM} 5.1.25(3))
\begin{equation}
\frac{m}{\Theta}\sum_{k=-\infty}^{\infty}\frac{1}{\left(  \frac{\lambda
}{\Theta}+2k\pi\right)  ^{2}+\left(  \frac{m}{\Theta}\right)  ^{2}}%
=\frac{\sinh\frac{m}{\Theta}}{\cosh\frac{m}{\Theta}-\cos\frac{\lambda}{\Theta
}},
\end{equation}
we get
\begin{equation}
\left\langle {\bar{\psi}\psi}\right\rangle =\sinh\frac{m}{\Theta}\times
\int_{0}^{2\pi}\frac{\rho\left(  \varphi\Theta\right)  }{\cosh\frac{m}{\Theta
}-\cos\varphi}d\varphi.\label{sdp}%
\end{equation}

So we come to the conclusion that $\left(  \ref{p}\right)  $ is nothing but a
particular case of $\left(  \ref{sdp}\right)  $ for the periodic spectral
density $\rho\left(  \lambda\right)  $ with the period $\Theta=T.$ It easy to
check that it leads to the expression for $\left\langle {\bar\psi\psi
}\right\rangle $ that coincides with $\left(  \ref{p}\right)  $ obtained in a
toy model approximation$.$

If one claims
\begin{equation}
\rho(\lambda)=\rho(\lambda_{1};\lambda_{2}\lambda_{3})=\rho(\lambda_{1}+2\pi
k_{1}T;\lambda_{2}+2\pi k_{2}T;\lambda_{3}+2\pi k_{2}T)\label{r3}%
\end{equation}
and
\begin{equation}
\sum_{\alpha=1}^{3}\lambda_{\alpha}=2\pi k_{0}T,
\end{equation}
where all $k_{j}$ are integer, one can easily get an expression for
$\left\langle {\bar\psi\psi}\right\rangle $ which coincides with the result
obtained in \cite{P} for $SU(3)$ gauge group case\footnote{Of course, specific
form of periodicity in $\left(  \ref{r3}\right)  $ was chosen intentionally to
achieve such coincidence.}. So we come to the conclusion that the 'toy' model
gives the results that may appear reasonable in the general case $\left(
\xi\sim1\right)  $ if periodicity assumption happens to be allowable.

\section{Some formal consequences of discrete symmetries}

A very formal expression of the symmetry $\left(  \ref{s1}\right)  ,$ $\left(
\ref{s2}\right)  $ can be found if we take into account the following. As it
is known the transformation under $CPT$\ reversal can be presented as a
combination of the unitary\ transformation\footnote{This invariance under
chiral rotation $\psi\rightarrow\exp\left\{  i\gamma_{5}\theta\right\}  \psi
$\ rotated by an angle $\theta=\pi$ is, in fact, the remnant of chiral
symmetry of the massless theory.} $\psi\left(  t,\mathbf{x}\right)
\rightarrow\gamma_{5}\psi\left(  t,\mathbf{x}\right)  $ and inversion
$\psi\left(  t,\mathbf{x}\right)  \rightarrow\psi\left(  -t,-\mathbf{x}%
\right)  $. Under unitary transformation as well as under inversion the
massless Dirac operator $\frak{D}$ only changes the sign. Therefore, the
result of the unitary transformation of $\frak{D}-m$ can be mimicked by the
formal substitution $m\rightarrow-m.$ It is easy to check that $\rho\left(
\lambda\right)  \leftrightarrow\rho\left(  -\lambda\right)  $ invariance also
leads to $m\leftrightarrow-m$ symmetry
\begin{equation}
\Sigma\left(  m\right)  =\left\langle {\bar\psi\psi}\right\rangle
=\int_{-\infty}^{\infty}\frac{\rho\left(  \lambda\right)  }{m-i\lambda
}d\lambda=\int_{-\infty}^{\infty}\frac{\rho\left(  -\lambda\right)
}{m-i\lambda}d\lambda=-\Sigma\left(  -m\right)  .\label{ps-s}%
\end{equation}

To be specific, we confine ourselves to the case of $SU\left(  2\right)  $
gauge group. Taking into account that
\begin{equation}
\frac1{\cosh x\pm\cos\frac\varphi2}=2\sum_{j=0}^{\infty}\left(  \mp1\right)
^{2j}\chi_{j}\left(  \varphi\right)  e^{-\left(  2j+1\right)  \left|
x\right|  },
\end{equation}
where $\chi_{j}\left(  \varphi\right)  =\frac{\sin\left(  2j+1\right)
\frac\varphi2}{\sin\frac\varphi2}$ are the characters of irreducible
representations of $SU\left(  2\right)  $ group, we can obtain from $\left(
\ref{p}\right)  $%
\begin{align}
\left\langle {\bar\psi\psi}\right\rangle  & =4n_{f}\sinh\frac mT\sum
_{j=0}^{\infty}e^{-\left(  2j+1\right)  \left|  \frac mT\right|  }\left(
-1\right)  ^{2j}\left\langle \chi_{j}\right\rangle +\label{den-ex}\\
& \ \ \ \ \ \ 4n_{f}\sinh\frac{m+2m_{r}}T\sum_{j=0}^{\infty}e^{-\left(
2j+1\right)  \left|  \frac{m+2m_{r}}T\right|  }\left(  -1\right)
^{2j}\left\langle \chi_{j}\right\rangle ,\nonumber
\end{align}
so the breakdown of the symmetry $\left(  \ref{s2}\right)  $ (and consequently
$\left(  \ref{ps-s}\right)  $) by the second term in $\left(  \ref{den-ex}%
\right)  $ for any $m_{r}\neq0$ became apparent. Moreover, $S_{F}$ posses an
undesirable invariance under the interchange
\begin{equation}
m\leftrightarrow-m-2m_{r}.\label{ui}%
\end{equation}
This invariance is inherent neither for Dirac operator nor for the spectrum
$\left(  \ref{den}\right)  .$ It the limit $a_{\tau}\rightarrow0$ such
symmetry is cognate to the invariance under the hopping parameter interchange:
$K\rightarrow-K,$ which plays an important role in the emergence of
conjectural Aoki phases, widely discussed in literature (see e.g.,
\cite{A,AKU,SS}).

Although the mirror fermion input does not disappear in $a_{\tau}\rightarrow0$
limit, almost all mirror terms exponentially vanish for $\frac{m_{r}}%
{T}\rightarrow\infty$. The only surviving term $j=0,$ however, introduces
finite contributions into $\left\langle {\bar{\psi}\psi}\right\rangle $ which
breaks $m\rightarrow-m$ symmetry. Let us consider a very special case
$m_{r}\rightarrow\infty$ , in other words, put exactly $r=1$ in the action
$\left(  \ref{sf}\right)  $ \textit{from the beginning.}

It is easy to see that in this case
\begin{equation}
\Delta_{x^{\prime},x}=\!\!\!\delta_{\mathbf{x}^{\prime}\mathbf{x}}\left(
U_{0}\left(  \mathbf{x},t\right)  \delta_{t,t^{\prime}-1}-\left(  ma_{\tau
}+1\right)  \delta_{t,t^{\prime}}\right)  \simeq\delta_{\mathbf{x}^{\prime
}\mathbf{x}}\left(  U_{0}\left(  \mathbf{x},t\right)  \delta_{t,t^{\prime}%
-1}-e^{a_{\tau}m}\delta_{t,t^{\prime}}\right)  ,\label{mf}%
\end{equation}
therefore
\begin{equation}
\det\Delta_{x^{\prime},x}=\prod_{\alpha}\left(  e^{\frac mT}+e^{i\varphi
_{\alpha}}\right)  ,\label{dd}%
\end{equation}
and
\begin{equation}
\exp\left\{  -S_{F}\right\}  =n_{f}\det\Delta\det\Delta^{\dagger}=n_{f}%
\prod_{\alpha}\left(  \cosh\frac mT+\cos\varphi_{\alpha}\right) \label{S-eff}%
\end{equation}
where $e^{i\varphi_{\alpha}}$\ are eigenvalues of the Polyakov matrix
$\Omega_{\alpha\beta}=\delta_{\alpha\beta}e^{i\varphi_{\alpha}}.$ In this case
(at least in zero order in $1/\xi$ ) mirror fermion term does not appear at
all and terms leading in $1/\xi$ are invariant under the change $m\rightarrow
-m$ $.$ The mirror fermion contributions in $\left(  \ref{dd}\right)  $are
shifted into the area of infinite $m_{r}$ and undesirable symmetry $\left(
\ref{ui}\right)  $ is expired.

It is interesting to note that, after 'harmless' substitution $\ 1+a_{\tau
}m\simeq e^{a_{\tau}m}$ the periodicity in $\lambda$ may appear in a general
case. Indeed, we may expect that in the limit $\left(  a_{\tau}\rightarrow
0;N_{\tau}\rightarrow\infty\right)  $ the mass variable will survive only in
combination $\left(  e^{a_{\tau}m}\right)  ^{N_{\tau}q}=e^{\frac{m}{T}q}$,
where $q$ is some dimensionless number. Thereby we get periodicity
$m\rightarrow m+2\pi iT/q$ . Taking into account that eigenvalues $\lambda$ of
the operator $\frak{D}=\gamma^{\nu}\left(  \partial_{\nu}-iA_{\nu}\right)  $
defined by $\det\left(  \frak{D}-i\lambda\right)  =0$ are related to
eigenvalues $\lambda_{m}$ of the operator $\frak{D}-m$ by $\lambda_{m}%
=\lambda+im$ we also get (as can be seen e.g., from $\left(  \ref{den}\right)
$) periodicity in $\lambda$%
\begin{equation}
\lambda\rightarrow\lambda+2\pi T/q.
\end{equation}

Recall that for finite $N_{\tau},a_{\tau}$ and $m_{r}$ one may write in a toy
model approximation \cite{P}
\begin{equation}
-S_{F}^{eff}=\sum_{\alpha}\ln\det\Delta_{\alpha}^{\dagger}\Delta_{\alpha}%
=\sum_{\alpha}\Xi_{\alpha}^{*}\Xi_{\alpha},
\end{equation}
where
\begin{equation}
\Xi_{\alpha}\left(  m^{\prime}\right)  =\cosh\left(  N_{\tau}%
\operatorname{arcsinh}\left(  m_{r}a_{\tau}+ma_{\tau}\right)  \right)
+\frac12e^{\frac{m_{r}}T}\Omega_{\alpha}+\frac12e^{-\frac{m_{r}}T}%
\Omega_{\alpha}^{*}.
\end{equation}
Thereby at finite $a_{\tau}$ we may expect only approximate periodicity in
$\lambda.$ That evidently differs from the case of $m_{r}\rightarrow\infty$ ,
which leads to $\left(  \ref{S-eff}\right)  .$

\section{Zero modes}

Among the non-perturbative properties of quarks, those associated with the
topology of the gauge fields occupy a special place. An important insight can
be gained about the topological content of gauge field configurations in
finite temperature QCD by computing the low lying modes of the Dirac operator.
In the continuum the Dirac operator of massless fermions in a smooth
background gauge field with non-zero topological charge has zero eigenvalues.
The corresponding eigenfunctions are chiral and the number of left- and
right-handed zero modes are related to the topological charge of the gauge
configuration: $n_{L}-n_{R}=\nu$ \cite{AS}. Besides, the study of the Dirac
operator spectrum near $\lambda=0$\ is of fundamental importance for the
understanding of chiral symmetry breaking in gauge theories.

Unfortunately, the lattice formulation also introduces artifacts and there is
no exact index theorem in this context. Instead, what is observed with the
actions currently used is the ``zero mode shift'' phenomenon \cite{Vi} (i.e.
modes with small but non-zero eigenvalue and not exactly chiral). Typical
gauge field configurations contributing to the path integral are not smooth
and, in addition, lattice regularization breaks some other conditions of the
index theorem as well: the topological charge of coarse gauge configurations
is not properly defined and Dirac operator breaks chiral symmetry in an
essential way \cite{HLN}. Unlike the continuum, there is no division of
lattice gauge fields into disconnected classes since lattice gauge fields are
characterized by group elements on the links of the lattice and all link
variables can be deformed to unity \cite{KLS}.

A noticeable progress in overcoming such difficulties was recently achieved .
In particular, as it is argued in \cite{HLN}, the lattice QCD in fixed point
formulation solves most of mentioned problems and in such context index
theorem seems to be valid on the lattice. The lattice version of index
theorem, which connects the small real eigenvalues of Wilson-Dirac operator
and the geometrical definition of topological charge, was suggested in
\cite{He}.

Reliable zero mode separation in numerical computations still remains a
serious problem \cite{KLS}, so we hope that even crude analytical models,
which allow to estimate $\rho(\lambda)$ behavior \ in the area of small
$\lambda$, may appear useful. Notwithstanding the abrupt fall-off and
vanishing of $\rho(\lambda)$\ at $\lambda=0$\ measured in \cite{V} may be a
finite volume effect \cite{S}, the present data par excellence do not confront
the suggestion that $\rho(0)$ is very small and compatible with zero. As it
was shown in \cite{S2} the spectral density decreases as $\lambda$ approaches
zero, and the larger is $n_{f}$ --- the larger is the effect. The scenario
when $\rho(0)$ hits zero at $n_{f}\geq5$ is quite probable.

Dirac spectrum should be particularly sensitive to the effects of quark loops
since fermion determinant for small quark mass strongly suppresses gauge
configurations with small Dirac eigenvalues. Convincing arguments that
spectral density disappears at $\lambda=0$\ are given in \cite{S2}. As it is
pointed out there, the weight in averaging involves a fermion determinant
factor
\begin{equation}
\left[  \mathrm{Det}(\frak{D}-m)\right]  ^{n_{f}}\ =\ \left[  m^{\left|
\nu\right|  }\prod_{\lambda_{n}>0}(\lambda_{n}^{2}+m^{2})\right]  ^{n_{f}%
}.\label{ind}%
\end{equation}
\ To be more specific we turn to the simple case of toy model approximation
where fermion part of action can be totally separated from the gluon one and
in accordance with $\left(  \ref{S-eff}\right)  $ incorporated into the
measure by interchanging $d\mu\lbrack\varphi_{\alpha}\left(  x\right)
]\rightarrow d\tilde{\mu}[\varphi_{\alpha}\left(  x\right)  ]$\ with
\begin{equation}
d\tilde{\mu}[\varphi_{\alpha}\left(  x\right)  ]=\prod_{\alpha=1}^{N}\left(
\cosh\frac{m}{T}+\cos\varphi_{\alpha}\left(  x\right)  \right)  ^{n_{f}}%
d\mu\lbrack\varphi_{\alpha}\left(  x\right)  ],\label{meas}%
\end{equation}
with
\begin{equation}
d\mu=\prod_{n>m}^{N}2\sin^{2}\frac{\varphi_{n}-\varphi_{m}}{2}\delta\left(
\sum_{k=1}^{N}\varphi_{k}\right)  \prod_{k=1}^{N}\frac{d\varphi_{k}}{2\pi};
\end{equation}
So the measure $\left(  \ref{meas}\right)  $ has a set of $n_{f}$-multiple
zeroes, which evidently corresponds to the zeroes of $\left(  \ref{ind}%
\right)  $. The effective action $S_{G}^{eff}=S_{G}^{eff}[\varphi_{\alpha
}\left(  x\right)  ]$\ for gluon field (e.g. the most popular $S_{G}^{eff}%
,$\ suggested in \cite{GK,O}) remains finite and scarcely able to influence
either multiplicity or locus of the zeroes. Zero modes constitute no exception
and are completely defined by the zeroes of $\left(  \ref{meas}\right)  $.

As it is easy to see from $\left(  \ref{mf}\right)  $, the eigenvalues of the
matrix $\Delta_{x^{\prime},x}$ , defined by $\det\left(  \Delta-i\lambda
\right)  =0,$ can be easily found from $\left(  \ref{dd}\right)  $ after the
simple shift $m\rightarrow m-i\lambda$%
\begin{equation}
\lambda=-im+T\left(  \varphi_{\alpha}\pm\left(  2n+1\right)  \pi\right)
\label{la}%
\end{equation}
and one can easily show that the eigenvalues of the matrix $\Delta_{x^{\prime
},x}^{\dagger}$ differs from $\left(  \ref{la}\right)  $ only by the sign of
the real part.

The expression $\left(  \ref{la}\right)  $ plainly shows that real modes
appear only in the limit $m\rightarrow0$ and are placed at $T\left(
\varphi_{\alpha}\pm\left(  2n+1\right)  \pi\right)  .$ It should also be noted
that zero modes created by the fields $\psi^{\left(  \pm\right)  }=\frac
{1\pm\gamma_{0}}2\psi$ are divided into pairs $\pm\lambda$ as it is for
$\psi^{\left(  L\right)  }=\frac{1-\gamma_{5}}2\psi$ and $\psi^{\left(
R\right)  }=\frac{1+\gamma_{5}}2\psi.$ Although the total numbers of the pairs
$n_{L}+n_{R}$ and $n_{+}+n_{-}$ are equal, the corresponding sets of
eigenvectors, of course, do not obligatory coincide, so $\nu=n_{L}-n_{R}$ and
$\nu^{\prime}=n_{+}-n_{-}$ may differ. Therefore, the topological nature of
$\nu^{\prime},$ if any, needs special consideration.

It is not incurious that the spectral density can be treated simply as a
result of integration of the $QCD$ action over all variables $\varphi_{x}$ but
one $\varphi$ or the integration over the surface with the fixed average
$\chi\left(  \varphi\right)  =\frac1V\sum_{x}^{V}\chi\left(  \varphi
_{x}\right)  .$ For example, in case of $SU(2)$ gauge group one may write
\begin{equation}
\rho\left(  \frac\varphi2\Theta\right)  =n_{f}\left\langle \delta\left(
\chi\left(  \varphi\right)  -\frac1V\sum\chi\left(  \varphi_{x}\right)
\right)  \right\rangle \frac{d\mu\left(  \varphi\right)  }{d\varphi}=n_{f}%
\exp\left\{  -S_{\left(  1\right)  }\left(  \varphi\right)  \right\}
\frac2\pi\sin^{2}\left(  \frac\varphi2\right)  .\label{r}%
\end{equation}

The relation between $\lambda$ and the phases of Polyakov loop eigenvalues
$\varphi_{\alpha}$ makes it obvious that $\lambda$ are not fixed numbers, the
dependence $\lambda$ on the parameters of the theory are defined not only by
the fermion part of action, but by the gluon part as well. In particular, zero
modes are defined by the relative number of Polyakov loops with $\cos
\varphi_{\alpha}$ $\simeq$ $-$ $\cosh\frac{m}{T}\simeq-1$ . It is obvious that
when the gauge fields are not fixed $\left(  \varphi_{\alpha}\left(
\mathbf{x}\right)  \neq const\right)  ,$ the behavior of the corresponding
$\det\Delta\left(  \varphi\right)  $ in the limit $m\rightarrow0$ is generally
different for different $\varphi_{\alpha}\left(  \mathbf{x}\right)
\mathbf{.}$ Therefore, $\left(  \ref{r}\right)  $does not result in an integer
value for\texttt{\ }index $\nu$ and we obtain the probability distribution,
$p(\nu)$, which is defined by $S_{\left(  1\right)  }\left(  \varphi\right)
$. However, we can compute the extreme value of $\nu:$ $\nu_{\max}$ $=\left|
\max\left\{  \nu\right\}  \right|  .$

Taking into account the condition $\sum_{\alpha}\varphi_{\alpha}=2\pi n$ we
get $\nu_{\max}=\left(  4+\frac2{n_{f}}\right)  \left[  \frac N2\right]  ,$
that both for $SU(2)$ and $SU(3)$ gives the same result $\nu_{\max}%
=4+\frac2{n_{f}}$.

In a series of papers \cite{SV,VZ,V,V2}, substantial new insight was added to
this issue. Their central assertion is that the spectral density of Dirac
operator very close to the origin $\lambda=0$ should be universal, depending
only on the symmetries in question. One of the amazing consequences of this
conjecture is that the spectral density of Dirac operator near the origin need
not be computed in full $QCD$, but can be extracted from simple models.

We cannot exclude that simple analytical structure of $S_{F}$ as in $\left(
\ref{S-eff}\right)  $\ will not persist in the general case, moreover, such
simplicity is in conflict with numerical results obtained in \cite{BDET}. If,
nevertheless, one ventures to apply universality arguments \cite{SV,VZ,V,V2}
to our model, it will lead to\texttt{\ }$\rho\left(  \lambda\right)
\sim\left(  \frac\lambda T\right)  ^{\nu}$ with $\nu=\nu\left(  g,T\right)  $
and $\nu_{\max}=4+\frac2{n_{f}}$. Thus, the configurations with small
eigenvalues are effectively suppressed, and the larger $n_{f}$\ is, the more
prominent is the suppression. Therefore, in the computation of averages their
contribution is small. It is true even for $\left\langle {\bar\psi\psi
}\right\rangle $ because ${\bar\psi\psi}$ has poles only in the points where
fermion determinant turns to zero.

\section{\texttt{\ }Conclusions}

We have presented some evidence in favor of periodicity suggestion for the
spectral density of Dirac operator on the lattice. Indeed, such periodicity
appears in the limit $a_{\tau}\rightarrow0$ in a toy model approximation
\cite{P} and in our opinion there are reasons to believe that it still
persists in the general case.

In this paper we take advantage of the fact that the chiral invariance of the
\textit{massless part} of Dirac operator leads to formal symmetry
$m\leftrightarrow-m.$ This symmetry is of importance in studying $QCD$ phase
structure in ($\mu;T;m$) space \cite{HJSSV}. It appears desirable to make such
symmetry explicit on finite size lattice. In case of infinitely heavy mirror
fermion $\left(  r=1\right)  $, the symmetry $m\leftrightarrow-m$ is restored.
Moreover, the periodicity of spectral density\ has also become explicit for
$r=1$ already at finite $a_{\tau}$.

Besides, we suggest a simple way to relate the spectral density $\rho\left(
\lambda\right)  $ to the quasiaverage $S^{\left(  1\right)  }$ of $QCD$ action
with Polyakov loop matrices with $\arg\Omega_{\alpha}\left(  \mathbf{\bar
x}\right)  =$ $\varphi_{\alpha}\left(  \mathbf{\bar x}\right)  $ at some point
$\mathbf{\bar x}$ being fixed. As it can be anticipated (see e.g.,\cite{ChCh})
the spectrum of Dirac operator is not invariant under $Z(N)$ transformations
and the results are different in the phases $\arg\Omega=0$\ and $\arg
\Omega=\frac{2\pi}3.$ Nonetheless, for $\arg\Omega=\frac{2\pi}3$ and
$\arg\Omega=-\frac{2\pi}3$ the results coincide, which reflects $C$ - symmetry
($\varphi_{\alpha}\leftrightarrow-\varphi_{\alpha}$) \cite{BPZ} at zero
density $\mu=0.$

In this paper, we have taken a modest first step towards analytical
investigation of Dirac spectral density characteristics. Apart from that there
are many other questions which can be answered with the low lying eigenvalues
and eigenvectors. The correlations between eigenvalues, the importance of
which was proved in \cite{KLS}, is one of them. However, it is much more
demanding a task than we can afford in such letter and we shall try to discuss
it elsewhere

\end{document}